\documentclass[twocolumn,aps,pra,floatfix,superscriptaddress]{revtex4-1}
\usepackage{amsfonts}
\usepackage{amssymb}
\usepackage{dcolumn}
\usepackage{graphicx}%
\usepackage{bm}
\usepackage{color}
\usepackage{braket}
\newcommand{\ambit}{\textsc{amb}{\footnotesize i}\textsc{t}}

\begin{document}

\title{High-precision \textit{ab initio} calculations of the spectrum of Lr$^+$}

\newcommand{\UNSW}{School of Physics, University of New South Wales, Sydney, New South Wales 2052, Australia
}
 
\newcommand{\HIM}{Helmholtz-Institut Mainz, 55128 Mainz, Germany}

\newcommand{\GSI}{GSI Helmholtzzentrum f\"ur Schwerionenforschung GmbH, 64291 Darmstadt, Germany}

\newcommand{\IFK}{Johannes Gutenberg-Universit\"at Mainz, Institut f\"ur Kernchemie, 55128 Mainz, Germany}

\newcommand{\TAU}{
School of Chemistry, Tel Aviv University, 6997801 Tel Aviv, Israel}

\newcommand{\RUG}{
Van Swinderen Institute for Particle Physics and Gravity,
University of Groningen, Nijenborgh 4, 9747 Groningen, The Netherlands\\
}

\author{E. V. Kahl}
\affiliation{\UNSW}

\author{J. C. Berengut}
\affiliation{\UNSW}

\author{M. Laatiaoui}
\affiliation{\IFK}
\affiliation{\HIM}

\author{E. Eliav}
\affiliation{\TAU}

\author{A. Borschevsky}
\affiliation{\RUG}

\date{\today}

\begin{abstract}
The planned measurement of optical resonances in singly-ionised lawrencium ($Z = 103$) requires accurate theoretical predictions to narrow the search window.
We present high-precision, \textit{ab initio} calculations of the electronic spectra of Lr$^+$ and its lighter homologue lutetium ($Z = 71$). We have employed the state-of-the-art relativistic Fock space coupled cluster approach and the \ambit\ CI+MBPT code to calculate atomic energy levels, g-factors, and transition amplitudes and branching-ratios. Our calculations are in close agreement with experimentally measured energy levels and transition strengths for the homologue Lu$^+$, and are well-converged for Lr$^{+}$, where we expect a similar level of accuracy. These results present the first large-scale, systematic calculations of Lr$^+$ and will serve to guide future experimental studies of this ion.

\end{abstract}

\maketitle

\section{Introduction}
The study of the transfermium elements (Z$>$100) lies at the frontier of contemporary nuclear and atomic physics research. The element synthesis itself provides a fertile terrain for studying effective interactions and nuclear matter under extreme conditions. Experimental shell gaps and single particle energies can be obtained from nuclear spectroscopy, which helps to improve model predictions for next spherical shell closures in the nuclear map: the location of the island of stability of superheavy elements.

Optical spectroscopy gives access to the atomic structure and provides insights into fundamental physics such as relativistic, correlation, and quantum electrodynamic (QED) effects. In addition, it can provide complementary information on single-particle and collective properties of atomic nuclei via hyperfine structure measurements. Such studies are continuously applied to ever heavier elements and are penetrating territories of the map that were previously inaccessible \cite{Laatiaoui19a}. A good example of this is the recent laser spectroscopy of the element nobelium ($Z=102$) \cite{Laatiaoui16a}, which demonstrated the technical feasibility despite a complete lack of tabulated spectral lines and production yields from nuclear fusion reactions of about one atom per second. These experiments have clearly shown how atomic modelling can efficiently support and guide atomic structure investigations and, in particular, that experiments and theory have to be pursued hand in hand. Current developments target the next heavier element, lawrencium ($Z=103$), in its neutral and singly charged states from both theory and experimental view points.

The planned experiments will attempt to optically excite Lr in a supersonic gas-jet: the Lr atoms are produced with high-energy from fusion reactions and are stopped and thermalised in a buffer gas cell. 
The gas-jet method enables to accelerate the lawrencium-buffer gas mixture into a low-pressure and low-temperature jet. This in turn allows to reduce collisional broadening and thus to increase the experimental resolution \cite{Laatiaoui19a, zadvornaya18a}. Previous experiments proved that the gas mixture contains both atomic species, neutral as well as singly ionised Lr, wherein the fraction of the ions substantially dominates the sample composition under typical experimental conditions~\cite{Kaleja19,Lautenschlaeger16}. For both species, due to the extremely low production yields, highly precise theoretical predictions of the spectral lines are required to develop excitation schemes and to narrow down the search window to be able to pinpoint the ground-state transitions. Moreover, predictions of lifetimes and branching ratios are needed to quantify experimental parameters such as required detector sensitivities and beam times.

In this work we provide high accuracy prediction of the energies and the $g$-factors of the low-lying excited states of Lr$^+$, along with transition rates and branching ratios between the different states. 
The calculations are performed within two complementary state-of-the-art relativistic approaches: the Fock space coupled 
cluster (FSCC) method \cite{EliKalIsh94, EliKalIsh94_2}, and the configuration 
interaction approach combined with many-body perturbation theory method (CI+MBPT) \cite{kahl19a}. In order to estimate the accuracy of our predictions for Lr$^+$, analogous calculations were performed for its lighter homologue, Lu$^+$, where we can compare the results of our 
calculations to experimental values.

While numerous predictions were reported for neutral Lr, to the best of our knowledge, no experimental and only three prior theoretical studies of atomic properties of Lr$^+$ are available. Dzuba \textit{et al} \cite{dzuba14a} calculated the first to the third ionization potentials of Lr using a linearized CI + all-order approach, while Cao and Dolg \cite{CaoDol03} calculated the first to the fourth ionization potentials of Lr using relativistic \textit{ab initio} pseudopotentials combined with the complete active space self-consistent field method and corrected for spin-orbit effects. In a much earlier publication, Fraga presented  a Hartree-Fock 
investigation of this system \cite{Fra74}; however, in that work, the ground state of Lr$^+$ was misidentified as $6d^2$. 

\section{Methods and Computational Details}

All the calculations were carried out within the framework of the projected Dirac-Coulomb-Breit Hamiltonian \cite{Suc80} (atomic units $\hbar = m_e = e = 1$ are used throughout this work),
\begin{eqnarray}
H_{DCB}= \displaystyle\sum\limits_{i}h_{D}(i)+\displaystyle\sum\limits_{i<j}(1/r_{ij}+B_{ij}).
\label{eqHdcb}
\end{eqnarray}
Here, $h_D$ is the one electron Dirac Hamiltonian,
\begin{eqnarray}
h_{D}(i)=c\, \boldsymbol \alpha_{i}\cdot \mathbf{p}_{i}+c^{2}(\beta _{i}-1)+V_\textrm{nuc}(i),
\label{eqHd}
\end{eqnarray}
where $\bm{\alpha}$ and $\beta$ are the four-dimensional Dirac matrices.  The nuclear potential $V_\textrm{nuc}$  takes into account the finite size of the nucleus.  The two-electron term includes the nonrelativistic electron repulsion and the frequency independent Breit operator,
\begin{eqnarray}
B_{ij}=-\frac{1}{2r_{ij}}[\boldsymbol \alpha_{i}\cdot \boldsymbol \alpha_{j}+(\boldsymbol \alpha_{i}\cdot \mathbf{r}_{ij})(\boldsymbol  \alpha_{j}\cdot \mathbf{%
r}_{ij})/r_{ij}^{2}],
\label{eqBij}
\end{eqnarray}
and is correct  to second order in the fine structure constant $\alpha$.

\subsection{FSCC}

We have calculated the transition energies of Lr$^+$ and its lighter homologue Lu$^+$ using the relativistic multireference valence universal
FSCC method, described in detail in Refs. \cite{EliKalIsh94, EliKalIsh94_2}. This approach is considered to be one of the most powerful methods for treatment of small heavy systems \cite{EliBorKal17}. Its particular advantage is the possibility of obtaining a large number of energy levels; it is therefore very well suited for calculating excitation spectra.

Our calculations start by solving the relativistic Hartree-Fock equations and correlating the closed-shell reference states for Lr$^{3+}$ and Lu$^{3+}$, which correspond to closed shell configurations. After the first stage of the calculation, two electrons were added, one at a time, to obtain the singly ionized atoms. At each stage of the calculations the appropriate coupled cluster equations were solved iteratively. To achieve optimal accuracy, large model spaces were used, going up to $13s11p9d8f6g5h$ for Lu$^{+}$ and $14s12p10d9f6g5h$ for Lr$^+$, and the convergence of transition energies with respect to the model space size was verified. In order to allow the use of such large model spaces without encountering convergence difficulties in the coupled cluster
iterations, the FSCC calculations were augmented by the
extrapolated intermediate Hamiltonian approach (XIH) \cite{EliVilIsh05}. 

The uncontracted universal basis set \cite{MalSilIsh93} was used, consisting of even-tempered Gaussian type orbitals, with exponents given by
\begin{eqnarray}
\xi _{n} &=&\gamma \delta ^{(n-1)},\text{ \ \ }\gamma =106\ 111\ 395.371\ 615 \\
\delta  &=&0.486\ 752\ 256\ 286. \nonumber
\label{eqUniversal}
\end{eqnarray}

The basis set used for both ions consists of 37 \textit{s} (\textit{n}=1--37), 31 \textit{p} (\textit{n=}5--35), 26 \textit{d} (\textit{n}=9--34), 21 \textit{f} (n=13--33), 16 \textit{g} (\textit{n}=17--32), 11 \textit{h} (\textit{n}=21--31), and 6 \textit{i} (\textit{n}=25--30) functions. The outer 60 electrons of Lu$^+$ and 74 electrons of Lr$^+$ were correlated, and virtual orbitals with energies over 200 a.u. were omitted. The FSCC calculations were performed using the Tel-Aviv Relativistic Atomic Fock Space coupled cluster code (TRAFS-3C), written by E. Eliav, U. Kaldor and Y. Ishikawa.

To account for the QED corrections to the transition energies we applied the model Lamb shift operator (MLSO) of Shabaev and co-workers
\cite{ShaTupYer15} to the atomic no-virtual-pair many-body DCB Hamiltonian as implemented into the QEDMOD program.
Our implementation of the MLSO formalism into the Tel Aviv atomic computational package
allows us to obtain the vacuum polarization and self energy contributions beyond the usual mean-field level, 
namely at the DCB-FSCCSD level.
 
\subsection{CI+MBPT}

Our calculations of the transition lifetimes and branching ratios, as well as the Land\`{e} $g$-factors for
the excited states of Lr$^+$ and Lu$^+$ were performed using the relativistic configuration 
interaction approach augmented with many-body perturbation theory method, via the \ambit\ atomic structure
software \cite{kahl19a}. We also present the transition energies calculated via this approach. The full details of this process have been extensively discussed elsewhere 
(see, for example Refs. \cite{kahl19a, dzuba96a, berengut16a, berengut06a, torretti17a, geddes18a}), so
we will only present a brief outline of the method here.

We start with a Dirac-Hartree-Fock calculation in the $V^{N-1}$ potential \cite{johnson94a};
that is, all but one electron
in the atom are included in the self-consistency calculations. This results in a set of Dirac-Fock
orbitals which are optimised for states with a single electron-excitation (i.e. $6s nl$ or $7s nl$ for
Lu$^+$ and Lr$^+$, respectively). Small-scale CI-only and CI+MBPT calculations showed that this choice
of potential produces closer agreement to experimental and FSCC energy levels than including all $N$
electrons in Dirac-Fock.

We generate a large basis of one-particle orbitals by diagonalising a set of B-splines over the one-electron Dirac-Fock operator \cite{johnson88a,beloy08a}.
We modify the operator to incorporate Lamb shift corrections via the
radiative potential method developed by Flambaum and Ginges \cite{FlaGin05}, which includes the
self-energy \cite{ginges16a} and vacuum polarisation \cite{ginges16b} contributions (finite 
nuclear-size effects are included using a Fermi distribution for nuclear charge). These corrections are propagated throughout the rest of the calculation by modification of the radial CI (Slater) and MBPT integrals.

Next, we use the B-spline basis functions to construct a set of many-electron configurations for the CI
expansion. We form the many-body functions by allowing all single and double excitations from the
$6s^2$/$7s^2$ ground-state up to $16spdfg$ (i.e. excitations with $n < 16$, and $0 < l < 4$). 
We then take
the Slater determinants with a given $M_J$ corresponding to these excitations and diagonalise the $J^2$ 
operator to form configuration state functions (CSFs), which are used to form the CI wavefunction via the 
standard CI eigenvalue problem \cite{berengut16a}.

We employ the emu CI method \cite{geddes18a, kahl19a} to significantly reduce the size 
of the CI eigenproblem by exploiting the fact that the CI expansion is typically dominated by 
contributions from $N_{\mathrm{dominant}}$ low-lying, dominant configurations. We divide the CI 
Hamiltonian matrix elements into three classes: leading diagonal elements; off-diagonal matrix 
elements containing at least one dominant configuration; and off-diagonal elements with no dominant 
configurations.
The contributions from the high-lying off-diagonal terms to the low-energy levels are small compared to the dominant terms, 
and so can be set to zero without significant loss of accuracy \cite{geddes18a, dzuba17a}. Typically 
$N_{\mathrm{dominant}} \ll N_{\mathrm{CI}}$, so
emu CI can significantly reduce the size of the CI matrix and thus computational load when compared to
standard CI. 

For both Lr$^+$ and Lu$^+$, we construct the dominant configurations from all single excitations up 
to $16spdfg$ and single and double excitations up to $12spdfg$; further increasing 
$N_{\mathrm{dominant}}$
changes the energy levels by less than $0.01\%$, suggesting this threshold captures all important
configurations. In both systems, increasing the basis size beyond $16spdfg$ changes the energy by 
$\sim 1\%$, indicating that the CI component of our calculations are well converged.

Additionally, we include corrections from core-valence correlations to second-order via the 
diagrammatic MBPT technique described in refs. \cite{dzuba96a, berengut06a}. We have included all 
one-, two- and three-body diagrams with orbitals up to $35spdfghi$ ($n \leq 35$, $0 \leq l \leq 6$). The MBPT 
corrections are rapidly convergent as more partial waves are added, and adding orbitals with $l \geq 
7$ to the MBPT basis changes the energy by less than $\sim 50$ cm$^{-1}$. Consequently, the MBPT 
component of our calculation is also well-converged.

The resulting CI+MBPT wavefunctions are used to calculate the Land\`{e} g-factors and
electric dipole transition matrix elements, which in turn give the transition lifetimes and branching
ratios.

For Lu$^+$ the experimental transition energy was used in the expression for Einstein coefficients, while for Lr$^+$ we used the calculated energies (our recommended values obtained from averaging the FSCC and the CI+MBPT results, see Section III for further details).

\section{Results}

Table \ref{TableI} contains the calculated ionization potential and transition energies of Lu$^+$, obtained with both approaches, along with the experimental energies. While many states are obtained in the calculations, here we present only the 8 lowest levels (from the $5d6s$ and the $6s6p$ configurations) that correspond to experimentally relevant transitions in Lr$^+$. Generally, the results are in good agreement with experimental values, with average differences between theory and experiment of $-263 \,(348)$~cm$^{-1}$ (where the number in brackets is the standard deviation of the difference) for the FSCC approach, and $16\,(389)$~cm$^{-1}$ for CI+MBPT. The two methods are also in good agreement with each other (average absolute difference of $278\,(496)$~cm$^{-1}$).
We expect similar accuracy for the calculated transition energies of the heavier homologue of Lu$^+$,  Lr$^+$, where no experimental data is yet available. 
The Breit interaction effect lowers the excitation energies by 20 -- 150~cm$^{-1}$, depending on the level. The QED corrections from both the MLSO formalism (for FSCC) and radiative potential method (for CI+MBPT) contribute between 100 --~200 cm$^{-1}$, also lowering the energies. Table \ref{TableI} also contains the calculated $g$-factors, which are overall in good agreement with experiment, indicating that the CI+MBPT approach successfully reproduces the character of the electronic wavefunction. A notable exception is the $g$-factor of the $^3P_2$ state, which is predicted to be 1.5, while the experimental value is reported as 1.66 \cite{MarZalHag78} (an assignment that may be erroneous).

\begin{table*}
 \caption{Ionization potential (top row), excitation energies, and $g$-factors of Lu$^+$ from CI+MBPT and FSCC calculations. Both results include the Breit and the QED corrections, the latter of which is shown in a separate column for comparison between the two calculations. Only levels relevant to the proposed Lr$^+$ experiment are presented here.}
  \centering
\begin{tabular}{clccccccc} 
\hline\hline 
&&	\multicolumn{2}{c}{$g$-factor} & \multicolumn{5}{c}{Energy (cm$^{-1}$)}  \\
\multicolumn{2}{c}{State}	&Exp. & {CI+MBPT} & {FSCC}  &{$\Delta$ QED} & CI+MBPT &{$\Delta$ QED}    &{Exp. \cite{MarZalHag78}} \\
\hline 
$6s^2$&$^1\!S_0$ IP	&--	&--  &112696    &-100    &--    & &111970 \\

5$d$6$s$ & $^3\!D_1$ &0.5	&0.52	&12354  &-158   &11664  &-144   &11796 \\
  & $^3\!D_2$        &1.16	&1.14 	&12985  &-156   &12380  &-143  &12432 \\ 
 & $^3\!D_3$         &1.33	&1.41	&14702  &-148   &14267  &-134  &14199 \\
  & $^1\!D_2$        &1.01	&1.09   &17892  &-157   &17875  &-160  &17332 \\ 
6$s$6$p$ & $^3\!P_0$ &--&--	        &27091  &-103   &27303  &-105  &27264 \\ 
     & $^3\!P_1$ &1.47	&1.51	    &28440  &-105   &28520  &-106  &28503 \\ 
  & $^3\!P_2$ &1.50	&1.66	        &32294  &-89    &32603  &-97  &32453 \\ 
  & $^1\!P_1$ &1.02	&0.99	        &38464  &-155   &37385  &-129  &38223 \\
\hline\hline %inserts single line
\end{tabular}
\label{TableI}
\end{table*}

\begin{figure}
    \centering
    \includegraphics[width=85mm]{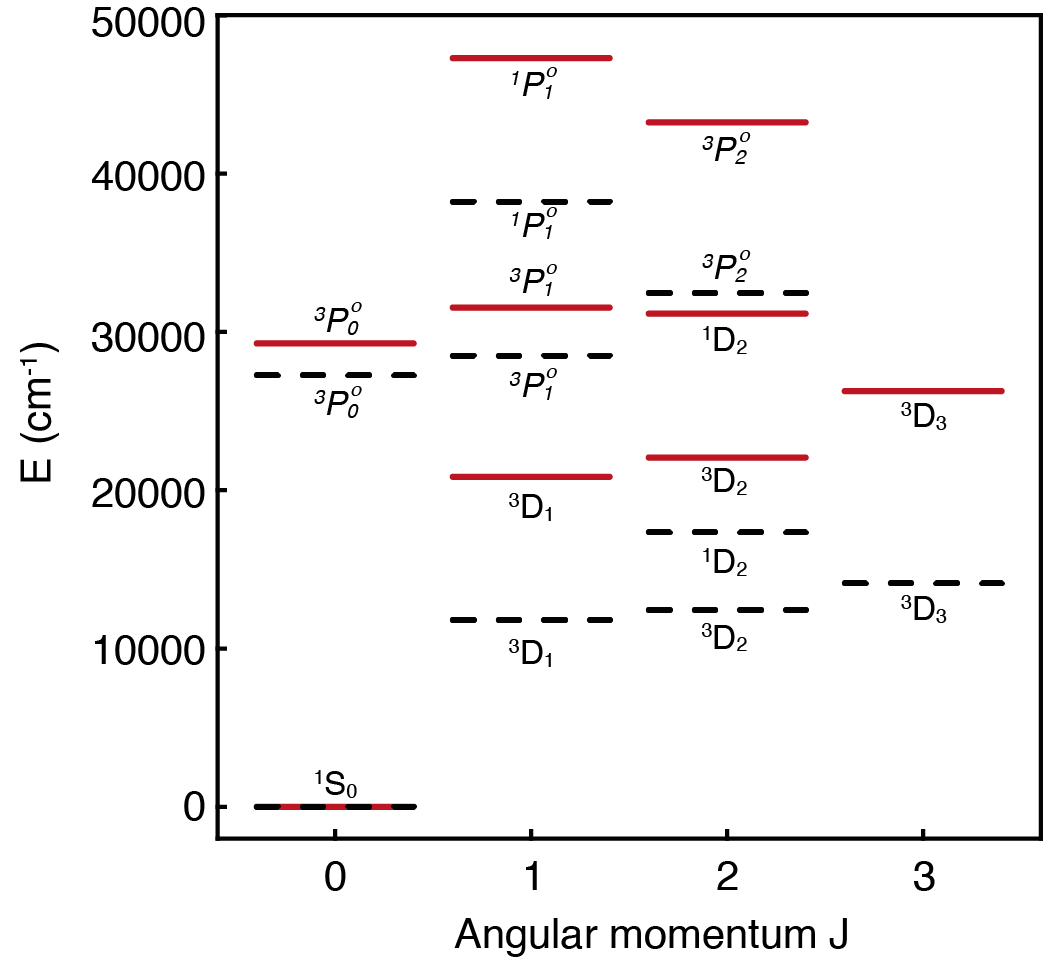}
    \caption{Grotrian diagram of experimental energy levels for Lu$^+$ (dashed, black) and recommended calculated energy levels for Lr$^+$ (solid, red). Levels are labeled by their approximate LS-coupling term symbol.}
    \label{fig:lr+_grotrian}
\end{figure}

Table \ref{TableII} contains the calculated ionization potential, excitation energies, and $g$-factors of the lawrencium ion. In all cases the energies are significantly higher than the corresponding levels in Lu$^+$ (see Grotrian energy-level diagram for both Lu$^+$ and Lr$^+$ in Figure 1). This is due to the relativistic stabilisation of the valence $7s$ shell in the heavier ion, which makes this system more inert.
The effect of the Breit interaction is higher in Lr$^+$ than in Lu$^+$, but the signs remain the same. Similarly, QED corrections in Lr$^+$ are slightly larger than in Lu$^+$, but remain on the order of 300~cm$^{-1}$, and are negative for all the considered states. The order of levels obtained using CI+MBPT and FSCC is the same, and the average difference between the two methods is $-47\,(747)$~cm$^{-1}$. 

Table \ref{TableII} also contains the recommended values for the excitation energies for the Lr$^+$ ion. The FSCC and CI+MBPT calculations have a comparable accuracy for Lu$^+$, often bracketing the experimental values. Consequently, our recommended transition energies are calculated as the mean of the FSCC and CI+MBPT results, with the conservative uncertainty estimate given by either the difference between the two calculated energies or the standard deviation of the difference between the CI+MBPT and experimental energy levels for Lu$^+$ (389~cm$^{-1}$), whichever is larger.

\begin{table*}
 \caption{Ionization potential (top row) and excitation energies of Lr$^+$ from CI+MBPT and FSCC calculations. Both results include the Breit and the QED corrections, the latter of which is shown in a separate column for comparison between the two calculations. The recommended values are obtained as the mean of FSCC and CI+MBPT results. Lifetimes and and $g$-factors derived from CI+MBPT calculations are also included.}
  \centering
\begin{tabular}{clccccccc} 
\hline\hline 
&&{$g$-factors}&	\multicolumn{5}{c} {Energies (cm$^{-1}$)} & Lifetime (s) \\
\multicolumn{2}{c}{State}&{CI+MBPT}	  &{FSCC}   &$\Delta \mathrm{QED}$ &{CI+MBPT}    &$\Delta \mathrm{QED}$    &Recommended  &   \\

\hline 
$7s^2$&$^1\!S_0$  & --	&116949     &-219  &--      &-- &116949 $\pm$ 389   &--\\

6$d$7$s$ & $^3\!D_1$ &0.5	&20265  &-342   &21426  &-374   &20846 $\pm$ 1161   &2.23 $\times 10^6$\\
  & $^3\!D_2$ &1.15	        &21623  &-344   &22507  &-373   &22065 $\pm$ 884    &8.26 $\times 10^{-2}$\\
  & $^3\!D_3$ &1.33	        &26210  &-326   &26313  &-355   &26262 $\pm$ 389    &2.97 $\times 10^{-2}$\\
  & $^1\!D_2$ &1.02	        &31200  &-373   &30942  &-397   &31071 $\pm$ 389    &1.53 $\times 10^{-3}$\\ 
7$s$7$p$ & $^3\!P_0$ &--	&29487  &-167   &29059  &-306   &29273 $\pm$ 428    &2.56 $\times 10^{-7}$\\ 
  & $^3\!P_1$ &1.42         &31610  &-179   &31470  &-314   &31540 $\pm$ 389    &1.45 $\times 10^{-8}$\\ 
  & $^3\!P_2$ &1.50	        &43513  &-240   &42860  &-308   &43186 $\pm$ 653    &2.43 $\times 10^{-8}$\\ 
  & $^1\!P_1$ &1.08	        &47819  &-260   &46771  &-376   &47295 $\pm$ 1048   &1.11 $\times 10^{-9}$\\
\hline\hline %inserts single line
 % is used to refer this table in the text
  \end{tabular}

  \label{TableII}
\end{table*}

Einstein $A$ coefficients (transition probabilities) for electric-dipole allowed (E1) transitions and  
branching-ratios for the transitions between the 8 lowest states in Lu$^+$ and for a number of other 
transitions where experimental results are available are shown in Table \ref{tab:Lu+_transitions}. 

Our calculated $A$ values are mostly larger than experimental values tabulated in 
\cite{sansonetti05a} by $~10\% - 30\%$, but the relative strengths are very well reproduced, and the strongest transitions are identified correctly.
The results of our CI+MBPT calculations for Lr$^+$ transitions are shown in Table 
\ref{tab:Lr+_transitions}. We expect a similar accuracy for the predicted Einstein coefficients and branching ratios to that obtained for the lighter homologue Lu$^+$. The $7s7p$ configurations can decay via electric dipole transitions, 
however the even-parity $6d7s$ states can only decay via $M1$ or $E2$ transitions to other even-parity states, for which the Einstein A coefficients are shown in Table \ref{tab:Lr+_M1+E2}.

The lifetimes of the Lr$^+$ levels, calculated via the Einstein A-coefficients, are presented in Table \ref{TableII}. Because $M1$ and $E2$ transitions are slow, even-parity states have significantly longer lifetimes than states which can decay via E1 transitions. In particular, the $6d7s\ ^3D_1$ state can only decay to the ground-state via a suppressed M1 transition, and so it has a lifetime of $2.2 \times 10^6$ seconds, or $\sim$25 days, which is several orders of magnitude longer than any of the other levels.

\begin{table*}
\caption{Einstein coefficients ($A_{\mathrm{CI+MBPT}}$) for dipole-allowed E1 transitions in Lu$^+$, calculated within the CI+MBPT approach using experimental transition energies, and compared to experimental values ($E_{\mathrm{NIST}}$, $A_{\mathrm{NIST}}$) where available \cite{sansonetti05a}. Note that levels which are not relevant to the proposed Lr$^+$ experiment and are not included in \cite{sansonetti05a} have been omitted, so branching ratios may not sum to 100\% for all levels.}

\begin{tabular}{l  l  l  l  l  l  l  l}
\hline
\multicolumn2{c}{Upper level} &\multicolumn{2}{c}{Lower level}  &$E_{\mathrm{NIST}}$
&$A_{\mathrm{CI+MBPT}} $(s$^{-1}$)	&$A_{\mathrm{NIST}}$ (s$^{-1}$)	&Branching ratio\\
\hline
\hline
%\multicolumn{6}{c}{}\\   
$6s6p$  &$^3P_0$ &$5s5d$ & $^3D_1$  &-- &2.19$\times 10^{7}$  &--  &1.00\\
\multicolumn{7}{c}{}\\
$6s6p$  &$^3P_1$ &$6s^2$ & $^1S_0$  &28503 &1.62$\times 10^{7}$  &1.25$\times 10^{7}$  &0.41\\
$6s6p$ &$^3P_1$ &$5d6s$ & $^3D_1$   &-- &6.84$\times 10^{6}$  &--  &0.17\\  
$6s6p$ &$^3P_1$ &$5d6s$ & $^3D_2$   &16707 &1.60$\times 10^{7}$  &9.90$\times 10^{6}$ &0.40\\  
$6s6p$ &$^3P_1$ &$5d6s$ & $^1D_2$   &-- &7.39$\times 10^4$  &-- &0.18$\times 10^{-3}$\\  
\multicolumn{7}{c}{}\\
$6s6p$ &$^3P_2$ &$5d6s$ & $^3D_1$   &-- &5.64$\times 10^{5}$  &--  &0.016\\  
$6s6p$ &$^3P_2$ &$5d6s$ & $^3D_2$   &-- &6.20$\times 10^6$  &-- &0.17\\  
$6s6p$ &$^3P_2$ &$5d6s$ & $^3D_3$   &-- &2.88$\times 10^7$  &-- &0.80\\
$6s6p$ &$^3P_2$ &$5d6s$ & $^1D_2$   &-- &3.56$\times 10^5$  &-- &9.91$\times 10^{-3}$\\
\multicolumn{7}{c}{}\\
$6s6p$  & $^1P_1$ &$6s^2$ & $^1S_0$ &38223 &5.21$\times 10^{8}$  &4.53$\times 10^{8}$  &0.96\\
$6s6p$  & $^1P_1$ &$6s^2$ & $^3D_1$ &-- &9.60$\times 10^{3}$  &--  &1.77$\times 10^{-5}$\\
$6s6p$  & $^1P_1$ &$6s^2$ & $^3D_2$ &-- &9.86$\times 10^{6}$  &--  &0.02\\
$6s6p$  & $^1P_1$ &$6s^2$ & $^1D_2$ &-- &1.07$\times 10^{7}$  &--  &0.02\\
\multicolumn{7}{c}{}\\
$5d6p$  & $^3D_1$ &$6s^2$ & $^1S_0$ &45532 &4.78$\times 10^{7}$  &7.14$\times 10^{7}$  &0.13\\
$5d6p$  & $^3D_3$ &$5d6s$ & $^3D_3$ &36298 &1.82$\times 10^{8}$  &1.66$\times 10^{8}$  &0.56\\
$5d6p$  & $^3D_3$ &$5d6s$ & $^3D_2$ &34534 &1.09$\times 10^{8}$  &9.20$\times 10^{7}$  &0.33\\
%$5d6p$  & $^3F_4$ &$5d6s$ & $^3D_3$ &34337 &2.70$\times 10^{8}$  &2.42$\times 10^{8}$  &0.95\\
\hline
\end{tabular}
\label{tab:Lu+_transitions}
\end{table*}

\begin{table*}
\caption{Einstein coefficients ($A_{\mathrm{CI+MBPT}}$) for dipole-allowed E1 transitions in Lr$^+$, calculated within the CI+MBPT approach and using our recommended calculated energies ($E_{\mathrm{calc}}$). Branching ratios for each transition are also 
shown.}
\begin{tabular}{l  l  l  l  l  l  l}                                           
\hline                                                                         
\multicolumn2{c}{Upper level} &\multicolumn{2}{c}{Lower level}  &$E_{\mathrm{calc}}$ 
(cm$^{-1}$) &$A_{\mathrm{CI+MBPT}} (s^{-1})$ &Branching ratio\\              
\hline                                                                         
\hline                                                                         
$7s7p$ &$^3P_0$ &$7s6d$  &$^3D_1$ &8515   &$5.44\times 10^6$  &1.00\\   
\multicolumn{7}{c}{}\\                                                         
$7s7p$ &$^3P_1$ &$7s^2$  &$^1S_0$ &31540  &$6.36\times 10^7$  &0.900\\          
$7s7p$ &$^3P_1$ &$7s6d$  &$^3D_1$ &10694  &$2.42\times 10^6$  &0.034\\
$7s7p$ &$^3P_1$ &$7s6d$  &$^3D_2$ &9475   &$4.66\times 10^6$  &0.066\\
$7s7p$ &$^3P_1$ &$7s6d$  &$^1D_2$ &608    &$4.54\times 10^{-1}$ &$6.4 \times 10^{-9}$\\
\multicolumn{7}{c}{}\\                                                         
$7s7p$ &$^3P_2$ &$7s6d$  &$^3D_1$ &22391  &$9.41\times 10^5$  &0.021\\          
$7s7p$ &$^3P_2$ &$7s6d$  &$^3D_2$ &21172  &$9.70\times 10^6$  &0.214\\          
$7s7p$ &$^3P_2$ &$7s6d$  &$^3D_3$ &16967  &$3.43\times 10^7$  &0.758\\          
$7s7p$ &$^3P_2$ &$7s6d$  &$^1D_2$ &12304  &$3.19\times 10^5$  &0.007\\ 
\multicolumn{7}{c}{}\\                                                         
$7s7p$ &$^1P_1$ &$7s^2$  &$^1S_0$ &47295  &$8.34\times 10^8$  &0.960\\          
$7s7p$ &$^1P_1$ &$7s6d$  &$^3D_1$ &26449  &$1.36\times 10^6$  &0.002\\         
$7s7p$ &$^1P_1$ &$7s6d$  &$^3D_2$ &25230  &$1.63\times 10^7$  &0.019\\          
$7s7p$ &$^1P_1$ &$7s6d$  &$^1D_2$ &16363  &$1.68\times 10^7$  &0.019\\ 
\hline
\end{tabular}
\label{tab:Lr+_transitions}
\end{table*}

\begin{table*}
\caption{Einstein coefficients ($A_{\mathrm{CI+MBPT}}$) for M1 and E2 transitions in Lr$^+$, calculated within the CI+MBPT approach and using our recommended calculated energies.}
\begin{tabular}{l  l  l  l  l  l  l}                                           
\hline                                                                         
\multicolumn2{c}{Upper level} &\multicolumn{2}{c}{Lower level}  &$E_{\mathrm{calc}}$ 
(cm$^{-1}$) &$A_{M1} (s^{-1})$ &$A_{E2} (s^{-1})$\\              
\hline                                                                         
\hline                                                                         
$7s6d$   &$^{3}D_1 $        & $7s^2$  &$^2S _0 $   &20846  &4.48 $\times 10^{-7}$ &--\\
\multicolumn{7}{c}{}\\                                                         
 $7s6d$  &$^3D _2 $     & $7s^2$  &$^2S _0 $   &22065  &-- &10.82\\
 $7s6d$  &$^3D _2 $     & $7s6d$  &$^3D _1 $   &1219   &0.79 &2.78 $\times 10^{-5}$\\
\multicolumn{7}{c}{}\\                                                         
 $7s6d$  &$^3D _3 $     & $7s6d$  &$^3D _1 $   &5416   &-- &0.0061\\  
 $7s6d$  &$^3D _3 $     & $7s6d$  &$^3D _2 $   &4197   &33.61    &0.015\\
\multicolumn{7}{c}{}\\                                                         
 $7s6d$  &$^1D _2 $     & $7s^2$  &$^2S _0 $   &31552  &-- &806.55\\
 $7s6d$  &$^1D _2 $     & $7s6d$  &$^3D _1 $   &10306  &49.81    &0.0694\\
 $7s6d$  &$^1D _2 $     & $7s6d$  &$^3D _2 $   &9087   &5.72 &0.0623\\
 $7s6d$  &$^1D _2 $     & $7s6d$  &$^3D _3 $   &4890   &5.67 &0.0034\\
\hline
\end{tabular}
\label{tab:Lr+_M1+E2}
\end{table*}

\section{Summary and conclusion}
We have calculated energies, $g$-factors, and lifetimes of several low-lying atomic levels in Lr$^+$. A striking agreement between the calculated FSCC and CI+MBPT energies is achieved. Similar calculations for the lighter homologue Lu$^+$ support the high accuracy of both approaches. 
In view of the prospects opened up by the forthcoming experiments, we identified two strong ground-state transitions in Lr$^+$, leading to $7s7p\,\,^3P_1$ and $7s7p\,\,^1P_1$ states at $31540\,$cm$^{-1}$ and $47295\,$cm$^{-1}$, respectively, that should in principle be amenable for experimental verification. 
In this case and for practical reasons, however, level searches are likely to focus on the $^3P_1$ state with a convenient transition wavelength and of smallest uncertainty.

\acknowledgements

The work of EVK was supported by the Australian Government Research Training Program scholarship. 
The authors would like to thank the Center for Information Technology of the University of Groningen for providing access to the Peregrine high performance
computing cluster and for their technical support.
We acknowledge Research Technology Services at UNSW Sydney for supporting this project with additional computing resources.
ML acknowledges funding from the European Research Council (ERC) under the European Union's Horizon 2020 research and innovation programme (grant agreement No. 819957). This work was supported by the Australian Research Council (Grant No. DP190100974).
AB is grateful for the support of the UNSW Gordon Godfrey fellowship.

\bibliography{Lr+}

\end{document}